\def\kms{km s$^{-1}$\ }

\documentstyle[12pt,emulateapj,amssym,flushrt,psfig]{article}

\lefthead{Padilla et. al}
\righthead{REDSHIFT-SPACE DISTORTIONS OF GROUP AND GALAXY CORRELATIONS}

\begin{document}

\title{Redshift-space distortions of group and galaxy correlations 
in the Updated Zwicky Catalog}

\author{Nelson D. Padilla, Manuel E. Merch\'an, Carlos A. Valotto and Diego G. Lambas}
\affil{nelsonp, manuel, val, dgl @oac.uncor.edu}
\affil{ Grupo de Investigaci\'on en Astronom\'{\i}a Te\'orica y Experimental (IATE), 
Observatorio Astron\'omico, Laprida 854, C\'ordoba and CONICET, Argentina. }
 
\and
 
\author{Marcio A. G. Maia}
\affil{ maia@on.br}
\affil{Departamento de Astromomia, Observat\'orio Nacional, Rua General 
Jos\'e Cristino 77, Rio de Janeiro, 20921-400, Brazil. }

\begin{abstract}

The two-point correlation function is computed for galaxies and groups of 
galaxies selected using 3-dimensional information from the Updated Zwicky Galaxy 
Catalog - (UZC).  The redshift space distortion of the correlation function 
$\xi(\sigma,\pi)$ in the directions parallel 
and perpendicular to the line of sight,
induced by pairwise group peculiar velocities is evaluated. 

Two methods are used to characterize the pairwise velocity field of groups and galaxies.  
The first 
method consists in fitting the observed
$\xi(\sigma,\pi)$ with a distorted model of an exponential 1-dimensional 
pairwise velocity distribution, in fixed $\sigma$ bins.  
The second method compares the contours of 
constant predicted correlation function of this model with the data.
The results are consistent with a 1-dimensional 
pairwise rms velocity dispersion of groups $<w^2>^{1/2}=250 \pm 110$ \kms.  
We find that UZC galaxy 1-dimensional pairwise rms velocity 
dispersion is $<w^2>^{1/2} = 460 \pm 35$ \kms.   Such findings point towards a 
smoothly varying peculiar velocity field from galaxies to systems of galaxies, as 
expected in a hierarchical scenario of structure formation.  

We find that the real-space correlation functions of galaxies and groups in UZC
can be well approximated by power laws of the form $\xi(r)=(r/r_0)^{\gamma}$. The 
values of $\gamma$ for each case are derived from the correlation function 
in projected separations $\omega(\sigma)$.  
Using these estimates we obtain $r_0$ from the projected correlation functions.
The best fitting parameters are 
$\gamma=-1.89 \pm 0.17$ and $r_0=9.7 \pm 4.5$ $h^{-1}$ Mpc for groups, and 
$\gamma=-2.00 \pm 0.03$ and $r_0=5.29 \pm 0.21$ $h^{-1}$ Mpc for galaxies.

The $\beta$-parameter ($\beta= \Omega^{0.6}/b$) is estimated for groups and galaxies 
using the linear approximation regime relating real and 
redshift-space correlation functions.  We find $\beta_{gx}=0.51 \pm 0.15$ 
for galaxies, in agreement with previous works, while for groups we obtain a 
noisy estimate $\beta < 1.5$.

Both methods used to characterize the pairwise velocity field are also 
tested on mock catalogs taken from CDM numerical simulations.
The results show that the conclusions derived from the application
of both methods to the observations are reliable.  
We also find that the second method, developed in this paper, 
provides more stable and precise results.

\end{abstract}

\section{INTRODUCTION}

The lack of homogeneity of the matter distribution 
on small scales, produces local
departures of the Hubble flow, inducing streaming motions of galaxies into the 
direction of larger concentrations of matter like 
clusters of galaxies or filaments.   The determination of the 
amplitude of peculiar velocities is important, 
since it is dependent on the cosmological matter density 
parameter, $\Omega$.  Thus, there 
is a possibility to constrain the value of $\Omega$ 
using the non-Hubble component of 
galaxy motions.   A characterization of this peculiar velocity field can be 
obtained by measuring the apparent distortion of the clustering pattern in the 
two-point correlation function for galaxies in 
redshift space, $\xi(\sigma,\pi)$,
where $\sigma$ and $\pi$ are the separations perpendicular and
parallel to the line of sight, respectively.  The anisotropy of the 
correlation function in redshift space depends on the peculiar velocity 
distribution function. On non-linear scales virialized regions
dominate, inducing the well known ``Finger of God'' effect, seen in 
redshift surveys allowing for
 estimates of the galaxy one-dimensional 
pairwise rms velocity dispersion $<w^2>^{1/2}$ (\cite{davis83}).  
On larger scales, where the linear regime applies, velocities are dominated
by the infall into overdense regions as bulk motion of galaxies
are generated by new levels of clustering.  This effect results in a 
compression of the $\xi$ contours
along the line of sight direction ($\pi$), 
and allows estimates of the parameter $\beta=
\Omega^{0.6}/b$ (\cite{kaiser87}), where $b$ is the linear bias parameter.
Extensive studies of $\xi(\sigma,\pi)$ for galaxies have been carried
out  (see for instance, \cite{loveday96}, 
\cite{ratcliffe98b}, \cite{tadros99}, and 
references therein).  The results of the different galaxy 
surveys are consistent 
with a one-dimensional pairwise velocity dispersion 
$<w^2>^{1/2} \simeq 400 \pm 50$ \kms, and parameter 
$\beta_{gx} \simeq 0.5 \pm 0.15$.

To examine larger volumes, clusters and groups of galaxies can be used 
as suitable tracers of the large scale structure of the universe.
Several works have 
characterized the clustering properties of rich clusters of 
galaxies (see for instance 
\cite{croft97} and \cite{abadi98}) concluding that the cluster autocorrelation 
function has a similar shape to that of galaxies, only with a larger
amplitude. An analogous conclusion was also obtained for groups of galaxies 
(e.g., \cite{merchan00}).  
Due to the fact that the amplitude of the correlation function 
of galaxy systems is 
significantly larger than that corresponding to galaxies, it is expected
a more significant contribution by the mean streaming velocity term.    
By contrast, galaxies with similar peculiar velocities would appear
strongly distorted along the line of sight due 
to their smaller correlation amplitude. 
Small scales effects, such as galaxy velocity 
bias and strongly non-linear regimes 
would not be present in the analysis of groups 
which could provide a more suitable 
estimate of the mean relative motions at 
intermediate and large scales.  

The subject of this paper is about redshift space distortions of galaxies
and groups of galaxies.  We study a sample of galaxies obtained from the 
Updated Zwicky Catalog (UZC, \cite{falco99}) and groups of 
galaxies (GUZC) obtained from this catalog  using a friends-of-friends 
algorithm (\cite{merchan00}).  The properties of these samples are briefly described in 
section 2.   In section 3, we estimate the pairwise 
characteristic velocity for groups applying a minimum 
$\chi^2$ analysis to the parameters of the exponential velocity distribution 
model. We also estimate the $\beta$ parameter. 
For comparison, we repeat the 
calculations for the galaxy parent catalog 
and compare our results with estimates from other works.  
In section 4, we test the adequacy of the methods applied to estimate 
$ <w^2>^{1/2} $ and $ \beta $ using mock catalogs taken 
from numerical simulations
corresponding to the empirical power spectrum 
(Peacock 1997) which reproduces the global
properties of galaxy clustering.
Conclusions are summarized in section 5.

\section{DATA}

The sample of galaxies 
was drawn from 
the UZC, containing 19,369 objects with apparent Zwicky magnitudes 
$m_{Zw} \le 15.5$ and 
with 96\% completeness in redshift. The region covered by the catalogue is
$20^h\leq \alpha_{1950}\leq 4^h$ and  $8^h\leq \alpha_{1950}\leq 17^h$
and $-2.5^o\leq \delta_{1950}\leq 50^o$, providing accurate coordinates 
within 2$"$ and reliable redshifts in the range $cz=0-25,000$ km s$^{-1}$ with a 
reasonably complete sky coverage 
(see \cite{falco99} for more details). For our statistical purposes, we have
considered galaxies with galactic latitudes $|b|> 20^o$ to avoid galactic obscuration.
Moreover, given the strong density gradient beyond $cz=15,000$ km s$^{-1}$ we adopted
this limit in $cz$.

The algorithm adopted by \cite{merchan00} 
for the construction of the catalog of groups 
of galaxies is basically the one described by \cite{huchra82} with the 
improvements by \cite{maia89} and by \cite{ramella97} in order to minimize the 
number of interlopers.  The groups present a surrounding density contrast  
($\delta\rho/\rho$),  relative to  the mean density of galaxies of 80.
The UZC group catalog (GUZC) contains systems with at least 4 members and 
mean radial velocities, $V_{gr} \le$ 15,000 \kms consistent with the
galaxy catalog. From the GUZC catalog 
we have considered groups with the same restrictions in galactic latitude 
and declination than those of the galaxy catalog
($|b|=20^o$ and $-2.5^o < \delta <50^o$).
Very rich groups containing more than 40 galaxies 
were not considered (only 6 groups) in the
sample to avoid few nearby large clusters and structures possibly related 
to systematics in the identification procedure biasing towards 
large percolating structures.  
In addition to the above criteria, those groups with $V_{gr} 
\le$ 2,000 \kms were
also discarded, to prevent any significant contribution of the group peculiar 
velocities in the measured redshifts 
which were used to determine their distances. 
With all the restrictions applied, the final sample is made up of 513 groups.

\section{ANALYSIS}

We analyze the effects of the peculiar velocity field on the 
correlation function of groups 
and galaxies as a function of the separations 
$\sigma$ and  $\pi$.
To compute $\xi(\sigma, \pi)$ we generate a random catalog with the 
same angular limits and radial selection function than the sample of objects.
We cross correlate data-data and random-random pairs ($N_{dd}$ and $N_{rr}$ 
respectively) binning them as a function
of separation in the two variables $\sigma$ and $\pi$.  Our estimate 
of $\xi(\sigma,\pi)$ is (Davis \& Peebles 1983):
\begin{equation}
\xi(\sigma,\pi)=\frac{N_{dd} n_R^2}{N_{rr} n_D^2}-1
\end{equation}
where $n_D$ and $n_R$ are the number of data and random points respectively. 

This estimator is sensitive to uncertainties in the mean density compared
to more stable estimators (eg. Hamilton 1993). However, since our analysis is
confined to the most inner scales where the correlation function has
a large amplitude, there is not a real need to use these alternatives.

The random catalog
has 400 times more objects than the data sample to minimize fluctuations.
To calculate $\xi(\sigma,\pi)$ for galaxies we have adopted the same 
cuts in radial velocity and galactic latitude
than for the sample of groups to make the results more comparable.
Within these restrictions, the final sample of galaxies comprises 14,755 objects.

In Figure 1 are displayed the levels of equal amplitude of the 2-point 
correlation function of groups (panel a) and 
galaxies (panel b) in the coordinates $\sigma$ and $\pi$. 
It can be appreciated the
compression of the groups iso-correlation curves in the $\pi$ direction which 
indicates the lack of high pairwise velocities 
in the sample of these systems of galaxies. 
From the comparison with clusters derived from surveys in two dimensions 
it is clear the existence of either large pairwise velocities or large 
projection biases as discussed by \cite{sutherland88}.
The presence of such projection biases in cluster 
samples derived in two dimensions 
has also been pointed out by \cite{vanHaarlem97}  
and by \cite{valotto00}. The sample of groups analyzed 
in this paper is derived from 
a nearly complete spectroscopic survey, so 
that we would not expect the artificial 
strong elongation along the line of sight observed 
in samples of clusters identified 
in two dimensions. This elongation would be originated by 
the systematic presence of groups along the line of sight in the fields
of 2-dim clusters.
From the theoretical point of view, such strong elongations along the line of
sight are not expected in a hierarchical scenario of structure formation.
This is confirmed by the compression observed in panel (a) and 
provides a clear evidence for the infall of groups onto larger structures. 
The level contours are well defined in the non-linear and mildly-linear 
regime $\xi(\sigma,\pi) \lesssim 1$.  The analysis 
developed in the next sections will therefore consider these level contours.
Due to the larger velocities of galaxies and their smaller 
correlation amplitude, it can be seen in panel (b) 
of this figure a large distortion of the 
iso-correlation curves in the $\pi$ direction.   
This is a direct consequence of the 
large relative peculiar velocities associated to the non-linear regime at small 
galaxy separations.

\subsection{Estimates of $<w^2>$}

In order to characterize the pairwise velocities of galaxies and 
groups in UZC from the correlation 
redshift-space distortion map, we have used two different 
procedures.  The first method corresponds to that adopted by 
\cite{loveday96} and \cite{ratcliffe98b}. 
We compare the observed correlation function  $\xi(\sigma,\pi)$ 
with the convolution of the real-space correlation function with the pairwise 
velocity distribution function $f(w)$ (\cite{bean83}  )

\begin{equation}
1+\xi(\sigma,\pi)=\int_{-\infty}^{\infty} [1+\xi(r)]f[w'+H_0 \beta \xi(r)
                  (1+\xi(r))^{-1} r']dw'
\label{pred}
\end{equation}

\noindent
where $r^2=r'^2+\sigma^2$, and $H_0$ is the Hubble constant, $r'=\pi - w'/H_0$ 
(the prime denotes the line-of-sight component of a vector quantity) and 
$<w> \simeq -H_0 \beta \xi(r) (1+\xi(r))^{-1}r'$ is the mean streaming velocity 
of galaxies at separation $r$.  We have calculated the best-fit rms peculiar velocity 
$<w^2>^{1/2}$ for an exponential distribution, 

\begin{equation}
f(w)=\frac{1}{\sqrt{2}<w^2>^{1/2}}\exp \left( -\sqrt{2} \frac{|w|}{<w^2>^{1/2}} \right) 
\label{fw}
\end{equation}

We adopted this pairwise velocity distribution because it has proved 
to be the most accurate fit to the results from numerical simulations as it has been 
shown by \cite{ratcliffe98b}.

We use a power-law model for $\xi(r)$ with an index $\gamma$ derived 
from the correlation function in projected separation $\omega(\sigma)$,  
$ \gamma = \gamma_p - 1 $, where $\gamma_p$ corresponds to the power law fit of 
the projected two-point correlation function.  
The projected correlations for groups and galaxies and the 
corresponding power-law 
fits ($\gamma_p=-0.89 \pm 0.17$ and $\gamma_p=-1.00 \pm 0.03$ for 
groups and galaxies 
respectively), are shown in Figure 2.

The optimum value of $<w^2>^{1/2}$ for this distribution was 
calculated for fixed 
$\sigma$ bins by minimizing the quantity $\chi^2$,

\begin{equation} 
\chi^2 = \sum_{\pi bins} [\xi^o(\sigma,\pi)-\xi^p(\sigma,\pi)]^2
\end{equation} 
 
\noindent where $\xi^o(\sigma,\pi)$ is the observed redshift space correlation 
function, and $\xi^p(\sigma,\pi)$ is the predicted correlation function from 
equation \ref{pred}.    The values of $\chi^2$ were calculated for different  
$r_0$ and $<w^2>^{1/2}$, and its minimum indicates the best pair of values 
$r_0$ and $<w^2>^{1/2}$.  The errors were obtained repeating the previous $\chi^2$ 
procedure for 50 bootstrap resamplings.   
The resulting values of $<w^2>^{1/2}$ for groups and 
galaxies are shown in tables 1 and 2.   The minimum $\chi^2$ values  are 
$<w^2>^{1/2}=390 \pm 185$ \kms and $<w^2>^{1/2}=520 \pm 95$ \kms for groups 
and galaxies respectively.

The second method adopted to characterize the pairwise velocity field compares the 
predicted and observed contours of constant correlation function.  We consider 
5 levels of constant correlation: $\xi(\sigma, \pi)$=$0.6$, $0.8$, $1.0$, 
$1.2$ and $1.4$.  
We describe the equal-correlation contours in polar coordinates $r$ and $\theta$
and we find the best fitting values of 
$r_0$ and  $<w^2>^{1/2}$  that minimize the quantity $\chi^2$ by summation
over the $\theta$ bins 

\begin{equation}
\chi^2 = \sum_{\theta bins}  [r^o(level,\theta)-r^p(level,\theta)]^2  
\end{equation}

\noindent where indexes $o$ and $p$ refer to the observed and predicted
radial coordinates of the correlation function contour levels, respectively. 

The standard deviation is again obtained 
from the dispersion of the distribution of 
best $r_0$-$<w^2>^{1/2}$ pairs for 50 bootstrap resamplings of the groups.
The results for groups and galaxies are shown in  table 3.
As it can be seen by inspection to this table,
this second method gives more stable determinations of $<w^2>^{1/2}$
(minimum $\chi^2$ values $<w^2>^{1/2}=250 \pm 110$ \kms and 
$<w^2>^{1/2}=460 \pm 35$ \kms for groups and galaxies respectively).
We note that the resulting value of $<w^2>^{1/2}$  for galaxies, is a factor 
$1.8$ higher than that of groups due to the high relative velocities of close 
pairs of galaxies.
 
\cite{dale99} estimate the 1-dimensional peculiar velocity dispersion for a sample 
of nearby Abell clusters $\sigma_v=341 \pm 93$ \kms.
This value can be compared with our estimates of $<w^2>^{1/2}$ that imply 
$\sigma_v^{gr} \lesssim 230$ \kms, considering the limiting case 
$<w^2> = 2 \sigma_v^2$ for statistically independent pairwise peculiar velocities.

We notice that neither method 1 nor method 2 provide a suitable constrain 
to the real space correlation length $r_0$.  In the next subsection we will 
estimate $r_0$ considering the projected correlation function, and use these 
values to infer the $\beta$ parameter using the linear theory approximation.

\subsection{Determination of the $\beta$-parameter}

In the linear regime, the direction-averaged redshift space 
correlation function, $\xi(s)$, and the real space correlation function, 
$\xi(r)$, are related by (\cite{kaiser87})

\begin{equation}
\xi(s) \simeq \left( 1 + \frac{2}{3} \beta + \frac{1}{5} \beta^2 \right) \xi(r)
\label{beta}
\end{equation}

We computed the redshift-space correlation functions for galaxies and groups, 
which are shown in Figure 3 for groups represented by a solid line and galaxies 
(dashed line). It can be seen from the inspection of this figure that both 
correlation functions can be well approximated by a power-law fit over a 
large range of separations. 

We define the projected 2-point correlation function, $W(\sigma)$, by

\begin{equation}
W(\sigma)=\int_{-\infty}^{\infty}\xi(\sigma,\pi)d\pi=
               2 \int_0^{\infty} \xi(\sigma,\pi) d\pi
\label{proy}
\end{equation}

The real-space correlation function $\xi(r)$ of groups and galaxies show 
larger negative slope values than $\xi(s)$.  For groups we obtain 
$\gamma_r \simeq -1.9$ and $\gamma_s \simeq -1.75$;  and for galaxies
$\gamma_r=-2.0$, $\gamma_s=-1.6$.  The difference between the two slopes is 
larger for galaxies than for groups due to their higher pairwise velocity dispersion.

To derive $r_0$ we use the definition of the Beta function in equation \ref{proy}, 
which gives

\begin{equation}
W(\sigma)= \sigma^{1-\gamma}r_0^{\gamma} \left[  \frac{\Gamma(\frac{1}{2}) 
                                          \Gamma(\frac{\gamma-1}{2})}
                                         {\Gamma(\frac{\gamma}{2})}    \right]
\label{r0}
\end{equation}
\noindent where $\Gamma(x)$ is the Gamma function and $\gamma<-1$ is assumed
(\cite{ratcliffe98a}).
We therefore find $r_0$ for different $\sigma$ bins, and using equation \ref{beta} 
we obtain the values of $\beta$.  In order to estimate the rms uncertainty in $\beta$, 
we estimate this parameter for 50 bootstrap resamplings to compute the dispersion 
around the mean. The results are shown in Figures 4 and 5.
We have estimated the real-space correlation length $r_0$ using the projected 
correlation function $W(\sigma)$ in equation \ref{r0}.
Errors are derived from bootstrap resamplings at different values of $\sigma$.  
It can be seen in Figure 4, the large scatter around the mean values 
for groups (panel a) and galaxies (panel b).   From a minimum $\chi^2$ analysis 
of the data points shown in both panels of Figure 4, we conclude that 
$r_0=9.7 \pm 4.5$  $h^{-1}$ Mpc  and $r_0=5.29 \pm 0.21$ $h^{-1}$ Mpc  
for $\sigma \in [1,12]$ $h^{-1}$ Mpc, 
are suitable estimates of the correlation length of groups and galaxies respectively.

The relative bias between optical galaxies and UZC groups can be derived from the 
real-space 2-point correlation functions.  Considering $r_0^{gx}=5.1$ $h^{-1}$ 
Mpc, $\gamma^{gx}= -1.6$ (\cite{ratcliffe98a}, this work), 
and our estimates for groups, 
$r_0^{gr}=8.6$ $h^{-1}$  Mpc, $\gamma^{gr}=-1.6$, 
we obtain $b^{gx}/b^{gr} \simeq 1.6$.

In Figure 5 we show the resulting $\beta$ parameter computed using equation \ref{beta}
for UZC groups (panel a), and for galaxies (panel b).  The
uncertainties in $\beta$ are obtained from propagation of errors 
by means of the standard deviation in $r_0$. 
We estimate a mean value of the $\beta$ using  a minimum $\chi^2$ 
statistics across the different 
$\sigma$ bins 
adopting $\Delta \chi^2=1$.  With a very large scatter, our results for 
$\beta_{gr}< 1.5$ do not provide a useful constraint.
The best $\beta$ value for galaxies ($\beta_{gx} =0.51 \pm 0.15$) is 
consistent with recent estimates from different surveys (\cite{ratcliffe98b}, 
\cite{tadros99}).

\section{TESTS WITH UZC MOCK CATALOGS}

To test the reliability of the results obtained in the previous sections we
have applied our statistical methods to mock catalogs extracted from N-body
simulations. The adopted model corresponds to the empirical power spectrum 
derived by \cite{peacock97} from the observed clustering of galaxies
corresponding to a flat universe with matter density parameter 
$\Omega=0.3$, cosmological constant density $\Omega_{\Lambda}=0.7$ 
and a Hubble constant of $H_0=65$ km/s Mpc$^{-1}$.
We use $128^3$ particles in a simulated box of $200$ $h^{-1}$ Mpc  
on a side normalized 
to $\sigma_8=0.76$. The mass per particle is 
$4.86 \times 10^{11}\ M_{\odot}$ so 
that in the following we will associate one galaxy to each particle.
From this box we extract a mock catalog of 51540 particles with the same 
radial selection function than 
UZC and $2.5$ times the area of the sky covered by UZC. 
We have also identified 1039 groups in the mock catalog with masses in the same
range of virial masses than in the observations using an equivalent 
procedure to that adopted by \cite{merchan00} 
The values of the one dimensional pairwise velocity dispersion is 
$<w^2>^{1/2}=400$ \kms for the mock groups and $<w^2>^{1/2}=520$ \kms  
for particles.
The correlation function of particles and groups are well fitted by power 
laws with $r_0=5.10 \pm 0.28 $ $h^{-1}$ Mpc, 
$\gamma=-1.50 \pm 0.08$
for particles and $r_0=8.46 \pm 0.47$ $h^{-1}$ Mpc, 
$\gamma=-1.54 \pm 0.11$ for groups. 
We show in Figure 6 the two-point correlation function of groups (panel a) and 
particles (panel b) in the space coordinates $\sigma$, $\pi$.
By inspection to this figure it can be appreciated the
compression of the groups iso-correlation curves in the $\pi$ direction similar 
to that shown in Figure 1.  On the other hand, the two-point correlation 
function of particles in $\sigma$, $\pi$ coordinates is strongly distorted along 
the $\pi$ direction due to the large pairwise particle velocities as well as 
their smaller correlation amplitude. 
These distortions are also similar to those 
shown in Figure 1 for the galaxy catalog.

We have computed the minimum $\chi^2$ value of the 1-dimensional pairwise rms velocity 
of groups and particles from
the comparison of the estimated (equation \ref{pred}) and the actual 
values of  $<w^2>^{1/2}$.
By application of the first method described in section 3.1, we obtain  
$<w^2>^{1/2}=140 \pm 100$ \kms and  $<w^2>^{1/2}=600 \pm 135$ \kms for 
groups and particles respectively, whereas the second method gives 
$<w^2>^{1/2}=430 \pm 85$ \kms and $<w^2>^{1/2}=530 \pm 70$ \kms.  
The results from the second method are in much better agreement with the 
actual values measured in the mock catalog than those from the first method 
(see table 3).   

We have also calculated the projected 2-point correlation function for groups 
and particles from the mock catalogs. Using equation \ref{r0} we obtain estimates 
of correlation length, $r_0=9.2 \pm 0.9$ $h^{-1}$ Mpc for groups, 
and $r_0=4.9 \pm 0.3$ $h^{-1}$ Mpc for particles (see Figure 7). 
These values agree very well with the ones of the computational box 
($r_0 = 8.46$ $h^{-1}$ Mpc and $r_0 = 5.1$ $h^{-1}$ Mpc for 
groups and particles, respectively), 
again indicating the efficiency of the methods adopted in deriving unbiased 
estimates of the parameter $\beta$.
In effect, by means of equation \ref{beta} and with the derived parameters of
the correlation function,
we obtain $\beta_{gr}<0.35$ (one standard deviation) and 
$\beta_{particles}=0.61 \pm 0.22$, which give a suitable agreement with the 
true values $\beta_{gr}=0.24$ and $\beta_{particles}=0.48$ respectively as
can be appreciated in Figure 8.

We notice the differences between mock and real data (Figures 4 and 5 compared
to Figures 7 and 8).  These are mainly due to the different number of
data points used, since we
aimed to test the statistical procedures with catalogs with similar
correlation and dynamical properties but with a larger number of points to improve
accuracy to test for small systematic effects.

Taking into account the success of the different tests carried out, 
we conclude that the methods applied in section 3 give reliable estimates 
of the real-space clustering and mean rms pairwise velocity of groups and 
galaxies providing suitable estimates of the $\beta$  parameter.
The tests also  indicate that the second method provides more stable and
confident results  of  $<w^2>^{1/2}$.

\section{CONCLUSIONS}

Two-point correlation functions of groups and galaxies were computed from 
the UZC.  The correlation function $\xi(\sigma,\pi)$ in the components 
parallel and perpendicular to the line of sight, allow us to characterize 
the pairwise peculiar velocities.   The distortions in $\xi(\sigma,\pi)$ 
are considered by the convolution of the real-space correlation function 
$\xi(r)$ with the pairwise velocity distribution 
$f(w)$ (eq. \ref{pred}).
We have adopted an exponential model according to eq. \ref{fw} and
we have used two methods to infer $<w^2>^{1/2}$.
The first method compares the model $\xi(\sigma,\pi)$ to the data in 
fixed $\sigma$ bins, while the second method compares contours of 
constant predicted and observed correlation function.  We use a minimum 
$\chi^2$ method to obtain the best fitting values of $<w^2>^{1/2}$.  More 
stable results come from the comparison of correlation 
level contours, where we obtain $<w^2>^{1/2}=325 \pm 175$ \kms
and $<w^2>^{1/2} = 433 \pm 42$ \kms for groups and galaxies respectively.  

We have estimated the real-space correlation function using a power-law fit 
$\xi(r)=(r/r_0)^{\gamma}$  for groups and galaxies.  
The correlation length, $r_0$, is obtained from the 
projected correlation function
$W(\sigma)$,
using the values of $\gamma$ derived from the correlation function in projected 
separations $\omega(\sigma)$.  
The best fitting values for the parameters are: $\gamma=-1.6$ 
and $r_0=8.6 \pm 3.1$ $h^{-1}$ Mpc for groups, and $\gamma=-1.8$ and 
$r_0=5.2 \pm 0.15$ $h^{-1}$ Mpc for galaxies. We have used these estimates of 
$r_0$ to compute $<w^2>^{1/2}$ for groups and galaxies.  The results for groups 
remain unchanged at $<w^2>_{gr}^{1/2} \sim 300$ \kms.  
For galaxies, the comparison at constant $\sigma$ bins gives values of 
$<w^2>_{gx}^{1/2}$ as low as $200$ \kms; on the other hand,
the comparison at constant $\xi$ levels 
has a better agreement with 
the results shown in section 4, $<w^2>^{1/2} \sim 500$ \kms, indicating the 
reliability of this procedure.

The results show a good agreement
well to those obtained in other optical surveys  
(e.g., \cite{loveday92}, \cite{loveday95}, \cite{lin96},  and 
\cite{ratcliffe98b}).   For instance, our value of
$<w^2>^{1/2}_{gx}=433 \pm 42$ \kms is entirely consistent with 
$<w^2>^{1/2}_{gx}=416 \pm 36$ \kms,  obtained for the Durham-UKST Galaxy 
Redshift Survey by \cite{ratcliffe98b}.  Our determination of the real-space 
correlation length $r_0=5.20 \pm 0.15$ $h^{-1}$ Mpc is also comparable to
that obtained in this survey $r_0= 5.1 \pm 0.3$ $h^{-1}$ Mpc  
(\cite{ratcliffe98a}),
although $\xi(r)$ is steeper in UZC ($\gamma = -1.8$) compared to 
$\gamma \simeq -1.6$ in the Durham-UKST redshift survey. 

We have also estimated the parameter $\beta= \Omega^{0.6}/b$ 
for groups and galaxies using the linear approximation relating real and 
redshift-space correlation functions.  Consistent with previous works for other 
optical samples (e.g., \cite{ratcliffe98b}, \cite{loveday96}) 
we find $\beta_{gx}=0.52 \pm 0.09$. 
Results for IRAS samples (Fisher et al. 1994, Tadros et al. 1999) are similar to those
derived for optical samples in spite of their
different bias parameters. As suggested by Fisher et al. there may be
a scale dependent galaxy biasing, although it should also be considered the possibility
that different methods sensitive to a variety of factors may provide different results
for a given sample.

For groups the results are noisy and do not allow a reliable estimate of
the $\beta$ parameter.  

The reliability of our results has been tested using mock catalogs extracted 
from N-body simulations and the identification of three dimensional groups in 
these catalogs. The correlation function of groups and particles in $\sigma$ 
and $\pi$ coordinates show a similar behavior than the one derived from 
the observations.

The estimated and real values of the dimensional pairwise rms velocity 
of groups and particles in the simulations are comparable showing that 
our procedure is capable of deriving reasonable estimates of 
$<w^2>^{1/2}_{gx}$ and $<w^2>^{1/2}_{gr}$.  Also, the best fitting 
parameters $r_0$ and $\gamma$ for the group and particle correlation 
function are in excellent agreement with the true values in the three 
dimensional box indicating that our methods provide good estimates of
the parameter $\beta$.

Given this fair agreement between derived and true values
we conclude that the small angle approximation
adopted in the analysis does not lead to systematic effects.
A similar argument holds for the derivation of the
$\beta$ parameter using \cite{kaiser87} linear formula
which gives a very good agreement with the actual value in the simulation,
supporting the reliability of the parameter $\beta$ derived from UZC galaxies.

Spurious large anisotropies along the line of sight obtained for Abell clusters
(e.g., \cite{sutherland91}) are significantly reduced
in the analysis of the APM cluster survey (\cite{daltonth}).
These large anisotropies would arise by inhomogeneities in the detection
of clusters in two dimensions and are entirely absent
in GUZC sample which shows the compression due 
to the infall onto larger structures.
The main reason why these anisotropies are not present in GUZC is 
the that the groups are identified in an homogeneous  catalog with angular
positions and redshifts.  The low value derived for the one-dimensional 
pairwise rms velocity of groups $<w^2>^{1/2} = 325 \pm 175$ \kms is indicative 
of a smoothly varying peculiar velocity field from galaxies to systems of 
galaxies, as expected in a hierarchical scenario of structure formation.  

We thank the Referee for helpful comments and suggestions which greatly improved
the previous version of this paper.
This work was partially supported by CONICET, SeCyT UNC and Fundaci\'on 
Antorchas, Argentina.  MAGM acknowledges CNPq grant 301366/86-1.

 {}

\newpage

\begin{deluxetable}{ccc}
\tablecolumns{3}
\tablenum{1}
\tablecaption{Pairwise Velocity Dispersion with First Method: Galaxies and particles }
\tablehead{
\colhead{ $\sigma [h^{-1} Mpc] $  }  &
\colhead{ $<w^2>_{gx}^{1/2} [km/s]$ } &
\colhead{ $<w^2>_{mock \ gx}^{1/2} [km/s] \ ^{(1)}$ } 
}
\startdata
 $0.5$&$430 \pm 59$&$600 \pm 122$ \nl
 $1.5$&$560 \pm106$&$500 \pm 117$ \nl
 $2.5$&$650 \pm112$&$890 \pm 164$ \nl
 $3.5$&$810 \pm140$&$560 \pm 121$ \nl
 $4.5$&            &$640 \pm 178$ \nl
 $5.5$&            &$500 \pm 194$ \nl
\enddata
\label{table1}
\tablenotetext{}{ $^{(1)}$ Actual value $<w^2>_{mock \ gx}^{1/2}=520\ km/s$.}
\end{deluxetable}

\begin{deluxetable}{ccc}
\tablecolumns{3}
\tablenum{2}
\tablecaption{Pairwise Velocity Dispersion with First Method: UZC and mock Groups }
\tablehead{
\colhead{ $\sigma [h^{-1} Mpc ] $  }  &
\colhead{ $<w^2>_{gr}^{1/2} [km/s]$ } &
\colhead{ $<w^2>_{mock \ gr}^{1/2} [km/s] \ ^{(1)}$ } 
}
\startdata
$1.5$&$460 \pm 198$&$410 \pm 101$ \nl
$4.5$&$360 \pm 189$&$ 50 \pm  89$ \nl
$7.5$&$360 \pm 170$&$ 10 \pm 153$ \nl
$10.5$&            &$ 30 \pm 105$ \nl
\enddata
\label{table2}
\tablenotetext{}{$^{(1)} $ actual value $<w^2>_{mock \ gr}^{1/2}=400\ km/s$}
\end{deluxetable}

\begin{deluxetable}{ccccc}
\tablecolumns{5}
\tablenum{3}
\tablecaption{Pairwise Velocity Dispersion with Second Method }
\tablehead{
\colhead{ $\xi $  }  &
\colhead{ $<w^2>_{gx}^{1/2} [km/s]$} &
\colhead{ $<w^2>_{mock \ gx}^{1/2} [km/s] \ ^{(1)}$ } &
\colhead{ $<w^2>_{gr}^{1/2} [km/s]$ } &
\colhead{ $<w^2>_{mock \ gr}^{1/2} [km/s] \ ^{(2)}$ } 
}
\startdata
 $0.6$&$430 \pm 43$&$500 \pm  82$&$340 \pm 156$&$400 \pm 185$ \nl
 $0.8$&$430 \pm 28$&$550 \pm  71$&$250 \pm 156$&$420 \pm 130$ \nl
 $1.0$&$480 \pm 39$&$470 \pm  92$&$240 \pm 140$&$410 \pm  95$ \nl
 $1.2$&$540 \pm 37$&$530 \pm  61$&$240 \pm  88$&$420 \pm  63$ \nl
 $1.4$&$430 \pm 34$&$560 \pm  50$&$240 \pm  85$&$450 \pm  64$ \nl
\enddata
\label{table3}
\tablenotetext{}{ \\
$^{(1)} $ actual value $<w^2>_{mock \ gx}^{1/2}=520\ km/s$. \\
$^{(2)} $ actual value $<w^2>_{mock \ gr}^{1/2}=400\ km/s$}
\end{deluxetable}

\begin{figure}[ht]
\centerline{\psfig{file=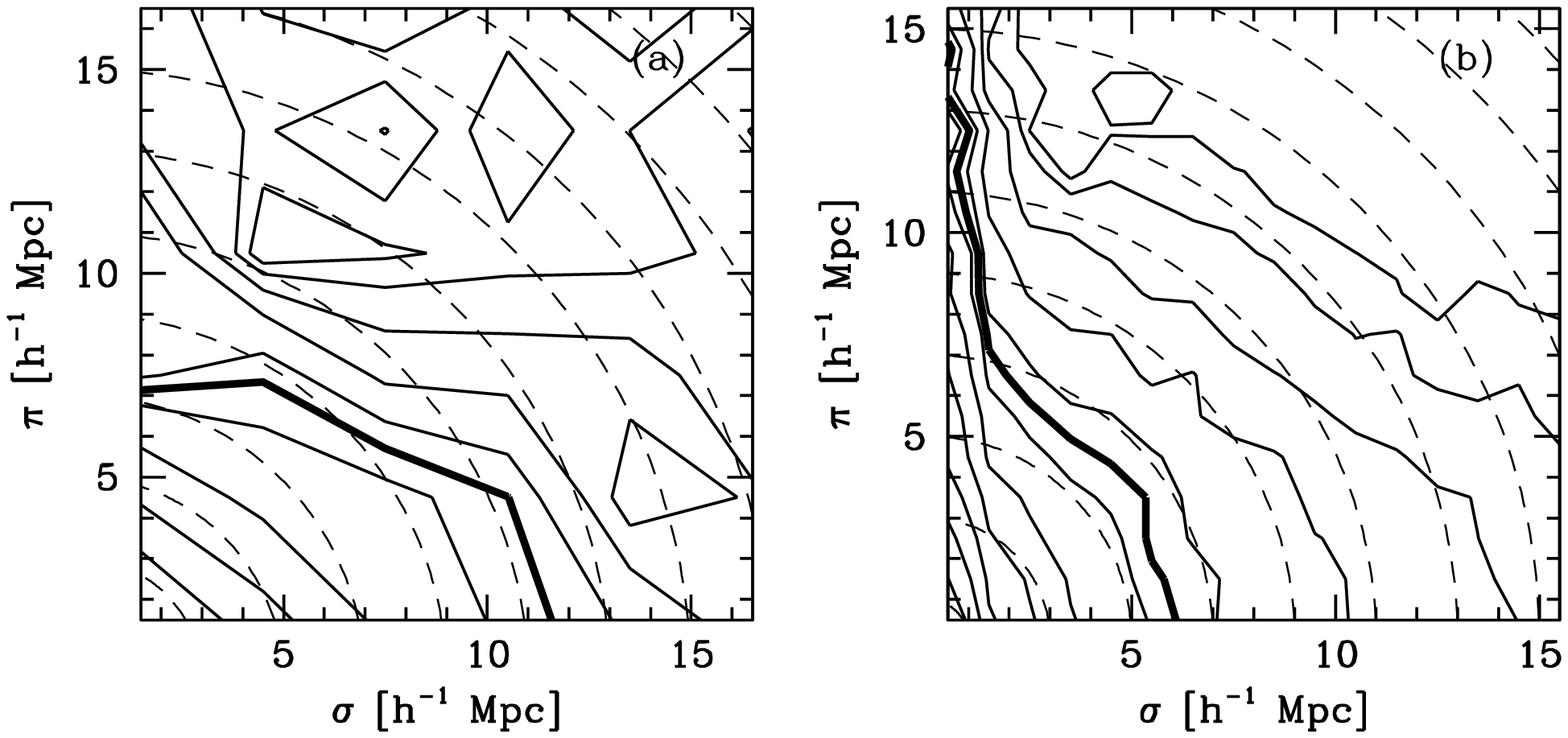,width=16cm}}
\caption{
Contour map of the correlation function $\xi(\sigma,\pi)$ estimated from
the UZC as a function of separations parallel ($\pi$) and perpendicular ($\sigma$) 
to the line of sight.  The contours correspond to fixed steps in $\log(\xi)$
from $-0.8$ to $1$.  The transition to the linear regime, $\xi(\sigma,\pi)=1$ level, 
is shown as a thick contour.  For comparison, dashed lines correspond to 
undistorted level contours. panel a) is for groups, while panel b) is for galaxies. }
\label{fig1}
\end{figure}

\begin{figure}[ht]
\centerline{\psfig{file=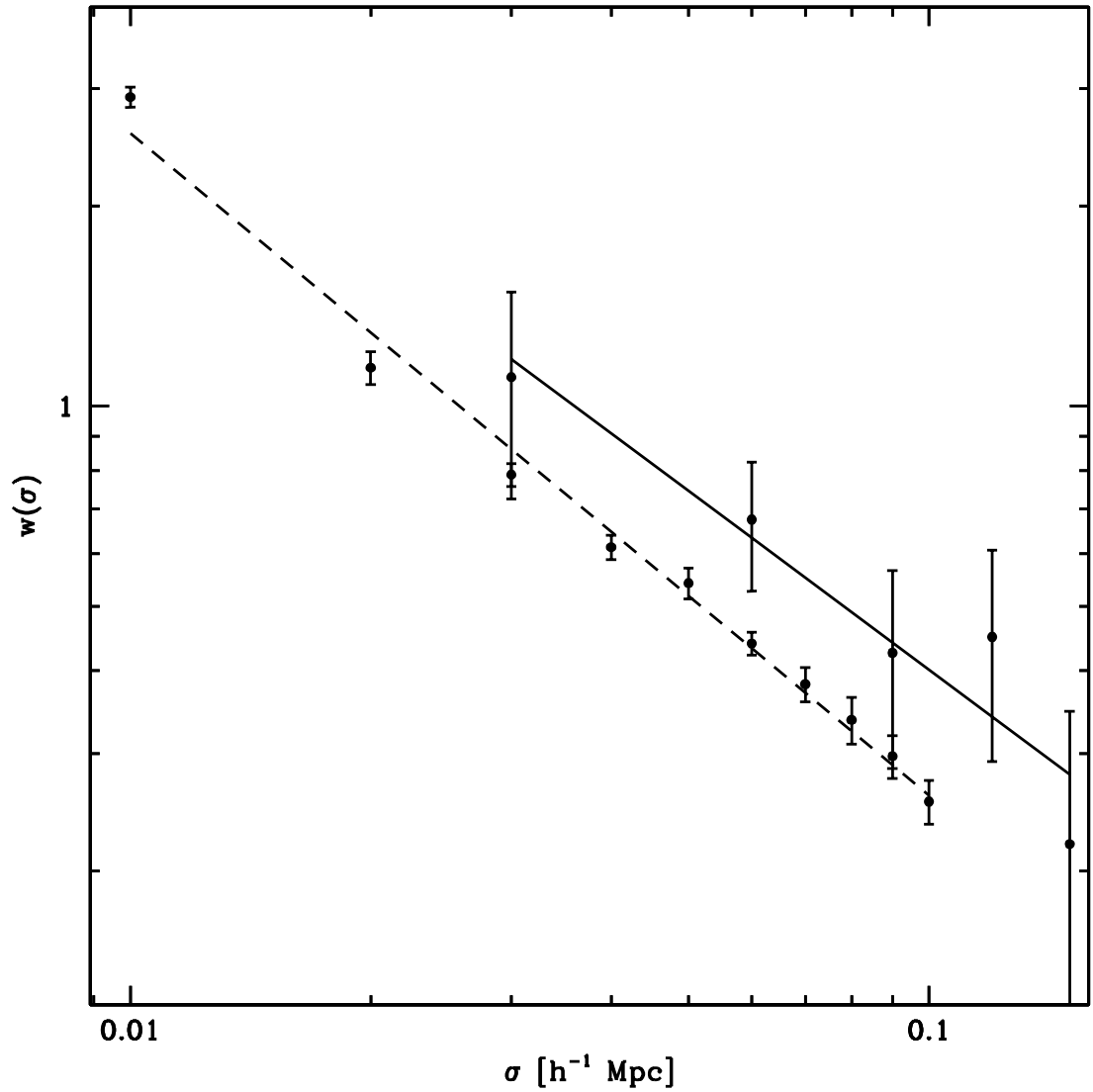,width=16cm}}
\caption{
Projected two-point correlation functions $\omega(\sigma)$ of UZC galaxies
and groups.  Solid and dashed lines correspond to the best power law fits
$A \sigma^{\gamma}$, with $\gamma=-0.89$ for groups and $\gamma=-1.0$ for galaxies
respectively.}
\label{fig2}
\end{figure}

\begin{figure}[ht]

\centerline{\psfig{file=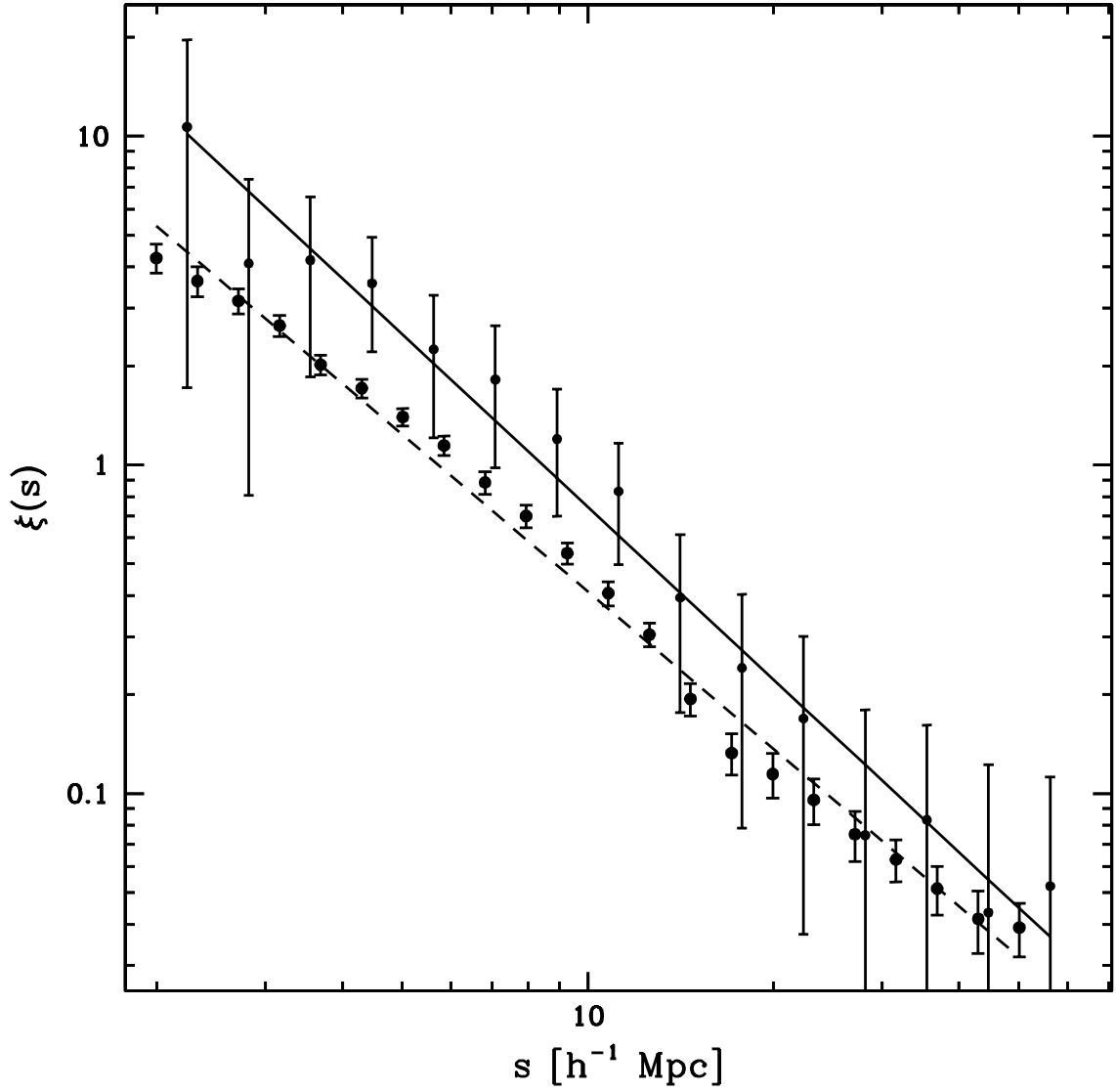,width=16cm}}

\caption{ Redshift-space autocorrelation function ($\xi(r)$) for galaxies and
groups. Error bars correspond to Poisson estimates of uncertainties. Solid and 
dashed lines show power law fits  $(r/r_0)^\gamma$ with $r_0=9.8$\ h$^{-1}$ Mpc,
$\gamma = -1.59 $ (groups) and  $r_0=6.4$\ h$^{-1}$ Mpc, $\gamma = -1.45 $ 
(galaxies) respectively.}
\label{fig3}
\end{figure}

\begin{figure}[ht]
\centerline{\psfig{file=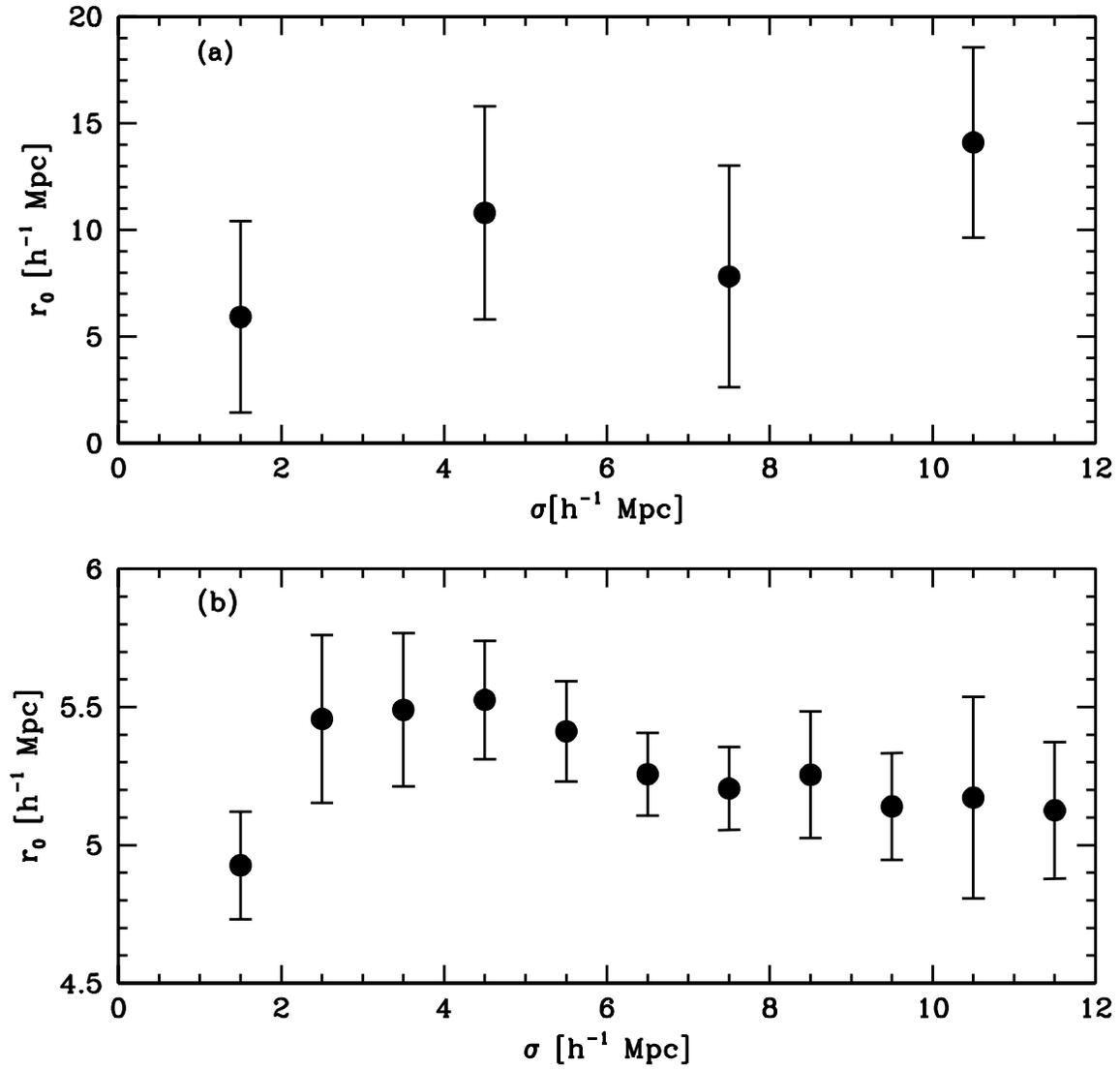,width=16cm}}
\caption{
Estimates of real-space correlation length $r_0$ derived  from the projected 
correlation function $W(\sigma)$ eq. \ref{r0}.  Errorbars correspond to direct 
estimates from bootstrap resamplings. a) UZC groups, b) UZC galaxies. }
\label{fig4}
\end{figure}

\begin{figure}[ht]
\centerline{\psfig{file=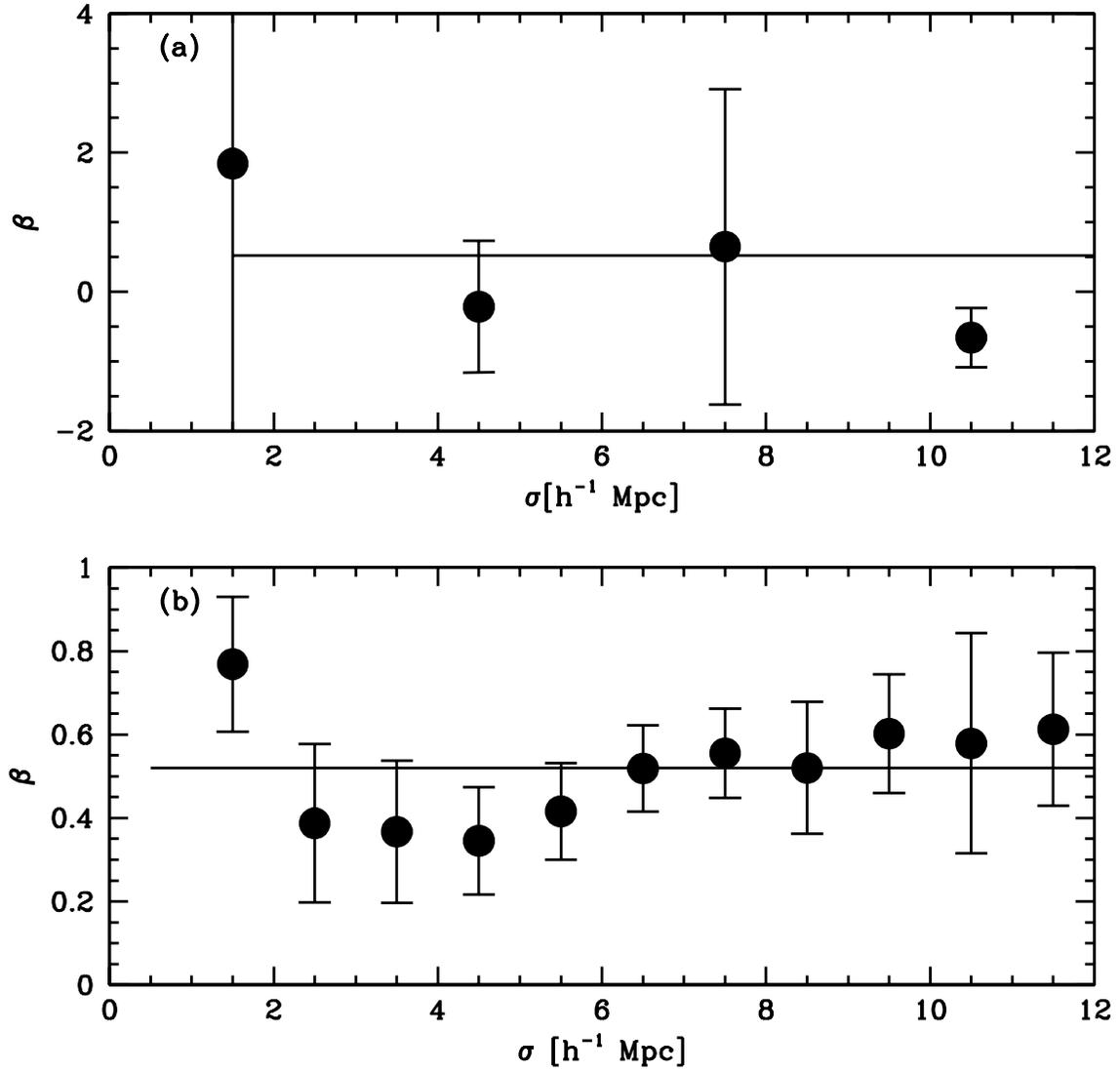,width=16cm}}
\caption{
Estimates of $\beta$ for UZC groups derived from equation \ref{beta}.
Errorbars correspond to propagation of errors in eq. \ref{beta} from errors
in $r_0$ quoted in figure 6.  The solid line corresponds to the 
best $\chi^2$ estimate of $\beta$ from galaxies.  a) UZC groups, b) UZC galaxies.}
\label{fig5}
\end{figure}

\begin{figure}[ht]
\centerline{\psfig{file=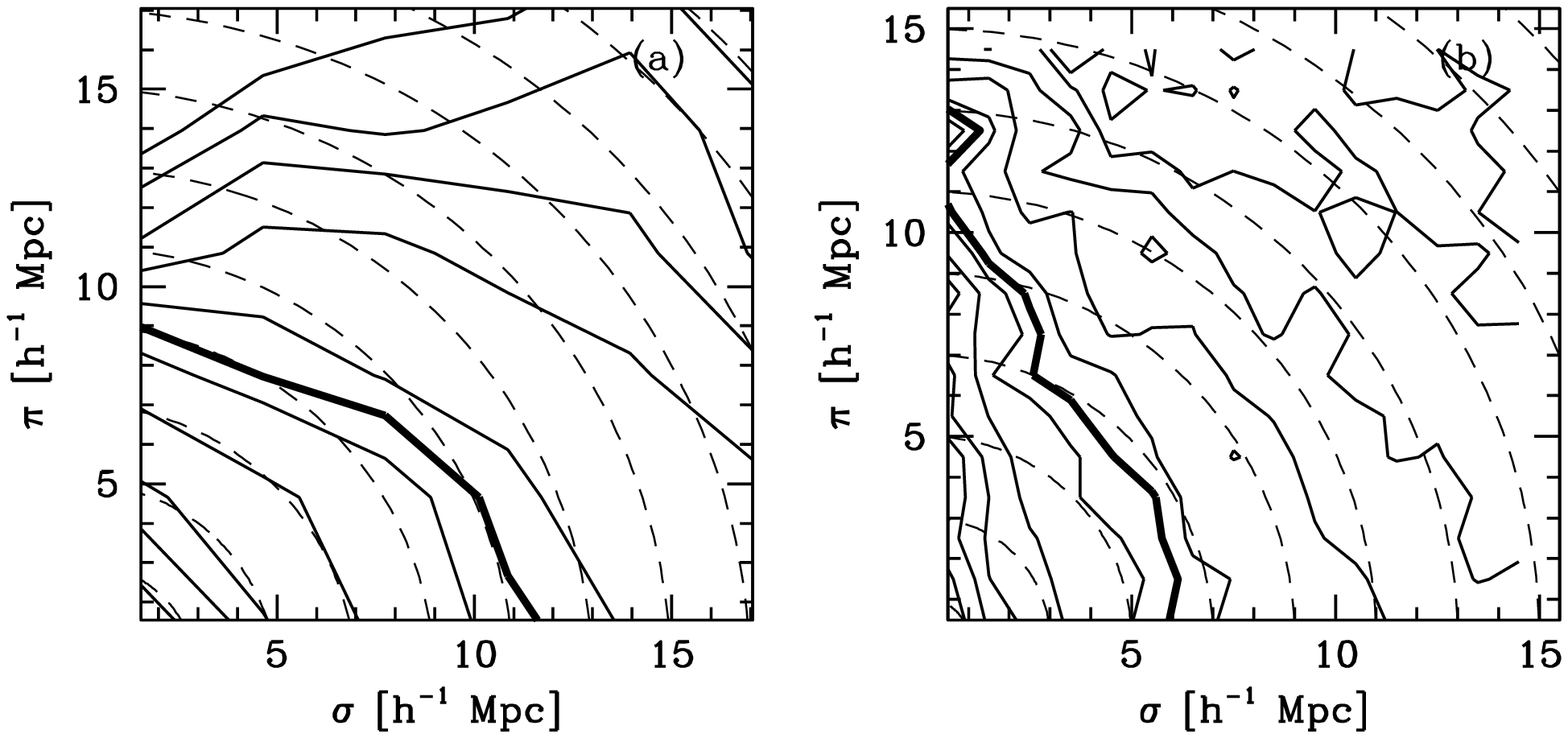,width=16cm}}
\caption{
Contour map of the correlation function $\xi(\sigma,\pi)$ estimated from
the mock catalogs as a function of separations parallel ($\pi$) and perpendicular 
($\sigma$) to the line of sight.  The contours correspond to fixed steps 
in $\log(\xi)$ from $-0.8$ to $1$.  
The transition to the linear regime, at $\xi(\sigma,\pi)=1$ level, 
is shown as a thick contour.  For comparison, dashed lines correspond to 
undistorted level contours. Panel a) is for mock groups, while panel b) is for mock 
galaxies.}
\label{fig6}
\end{figure}

\begin{figure}[ht]
\centerline{\psfig{file=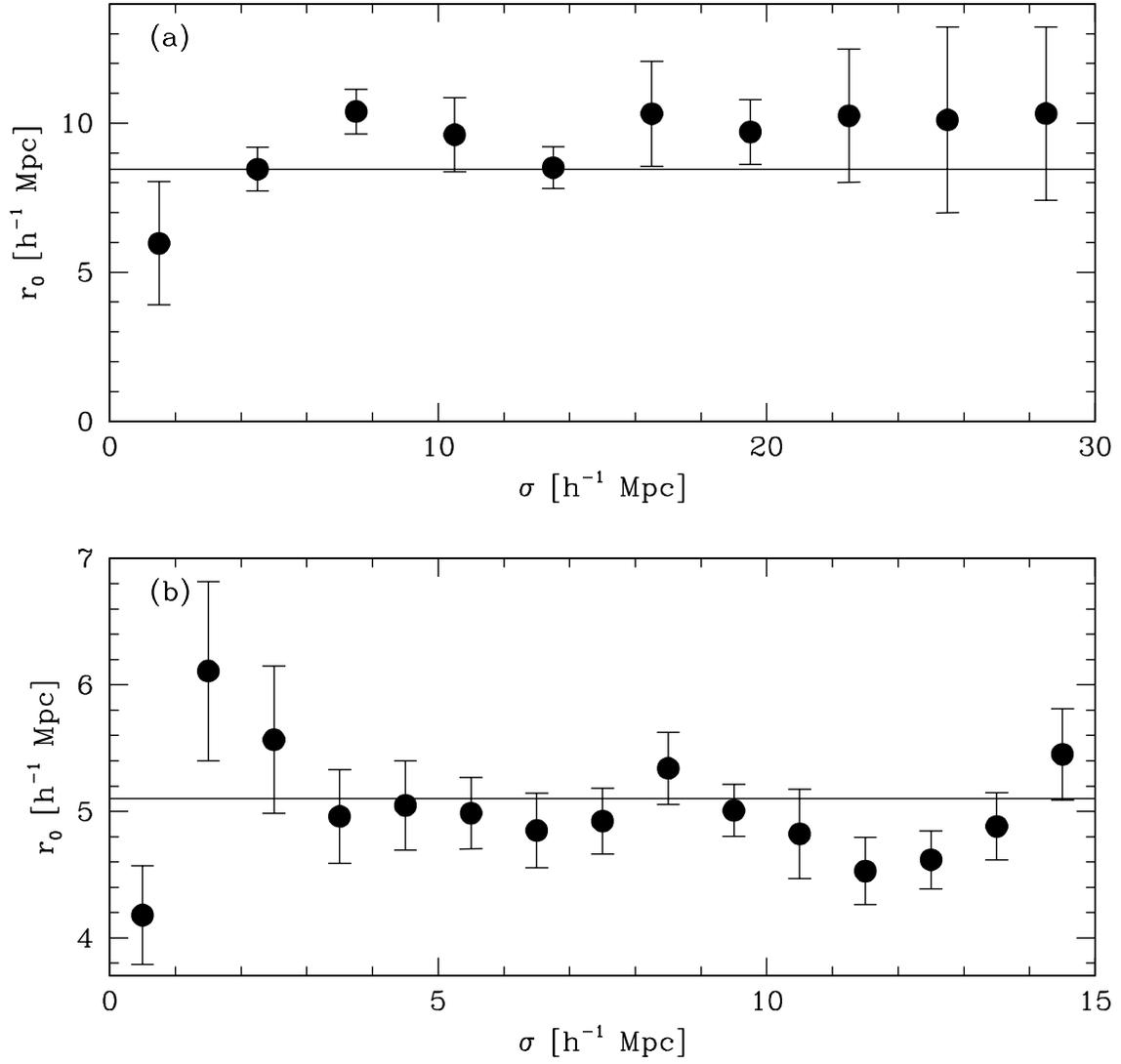,width=16cm}}
\caption{
Estimates of real-space correlation length $r_0$ derived  from the projected 
correlation function $W(\sigma)$ eq. \ref{r0}.  
a) mock groups, b) mock galaxies.
Errorbars correspond to direct 
estimates from bootstrap resamplings. 
The solid line corresponds to the 
true values of $r_0$ of galaxies and groups in the numerical simulation.  }
\label{fig7}
\end{figure}

\begin{figure}[ht]
\centerline{\psfig{file=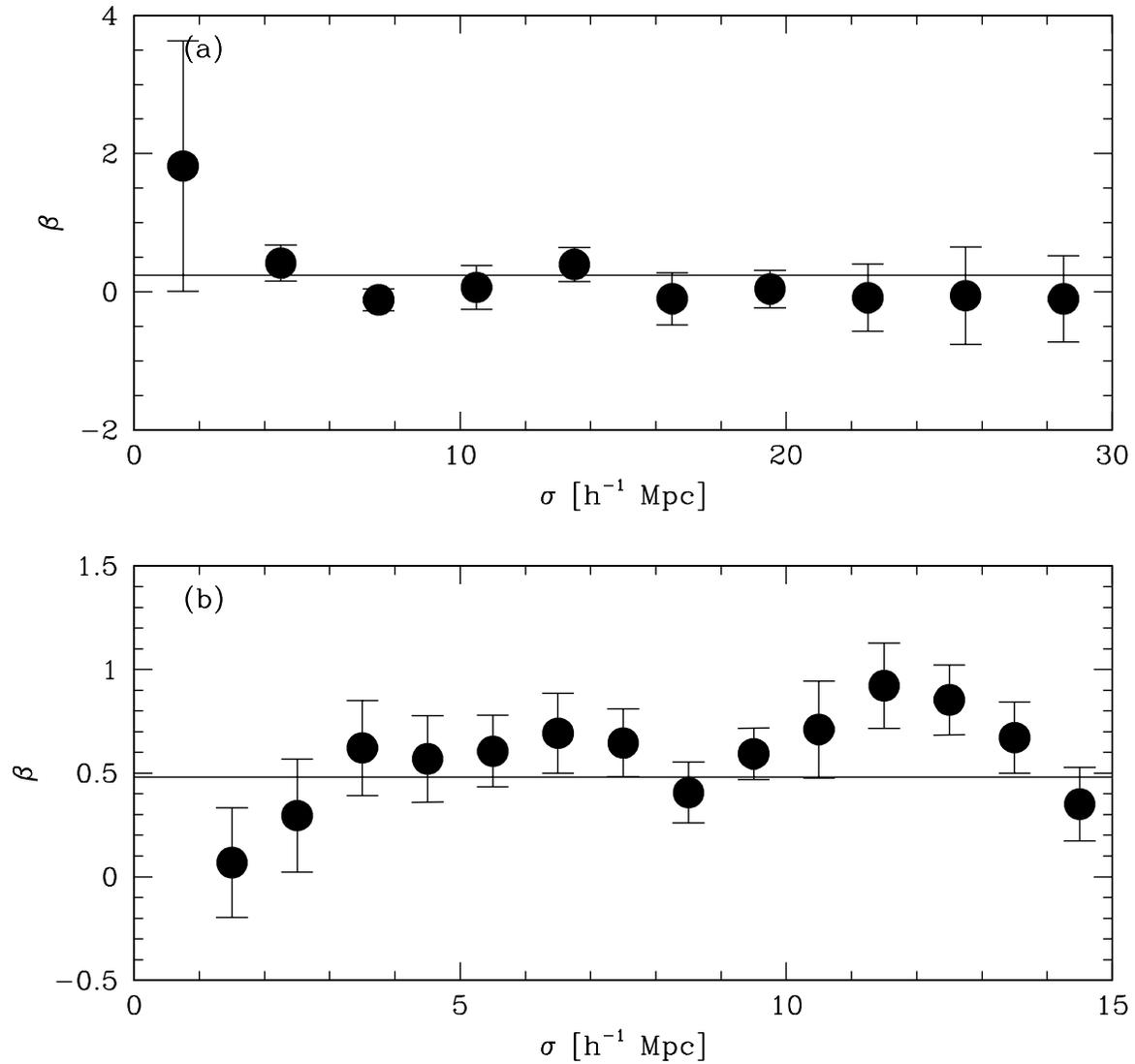,width=16cm}}
\caption{
Estimates of $\beta$ for mock groups derived from equation \ref{beta}.
a) mock groups, b) mock galaxies.
Errorbars correspond to propagation of errors in eq. \ref{beta} from errors
in $r_0$ quoted in figure 6.  
The solid line corresponds to the 
true values of $\beta$ of galaxies and groups in the numerical simulation.  }
\label{fig8}
\end{figure}

\end{document}